\documentclass[useAMS,usenatbib]{mn2e}
\usepackage{graphics}
\usepackage{amssymb}
\usepackage{amsfonts}
\usepackage{wasysym}

\newcommand{\s}{\scriptsize}
\newcommand{\ty}{\tiny}
\newcommand{\m}{\mathrm}

\title[Searching for CU\,Vir-type cyclotron maser from $\sigma$\,Ori\,E]
{Searching for CU\,Vir-type cyclotron maser from $\sigma$\,Ori\,E:\\
The role of the magnetic quadrupole component}
\author[P. Leto et al.]
{P. Leto$^{1}$ \thanks{E-mail: pleto@oact.inaf.it},
C. Trigilio$^{1}$,
C. S. Buemi$^{1}$,
F. Leone$^{2}$,
G. Umana$^{1}$\\
$^{1}$INAF - Osservatorio Astrofisico di Catania, Via S. Sofia 78, 95123 Catania, Italy\\
$^{2}$Universit\`a  di Catania, Dipartimento di Fisica e Astronomia, Sezione Astrofisica, Via S. Sofia 78, 95123 Catania, Italy}
\begin{document}

\date{}

\pagerange{\pageref{firstpage}--\pageref{lastpage}} \pubyear{}

\maketitle

\label{firstpage}

\begin{abstract}
{ In this paper we present new and archive radio measurements obtained with the Very Large Array
of the magnetic chemically peculiar (MCP) 
star $\sigma$\,Ori\,E. The radio data have been obtained
at different frequencies and are well distributed along the rotational phases.
We analyze in detail the radio emission from $\sigma$\,Ori\,E with the aim to search evidence of circularly polarized
radio pulses. Up to now, among the MCP stars only CU\,Virginis shows 100\% polarized 
time-stable radio pulses, explained as highly directive electron cyclotron maser emission, visible from Earth
at particular rotational phases, like a pulsar.}
Our analysis shows that
there is no hint of coherent emission at frequencies below 15 GHz.
We conclude that the presence of a quadrupolar component of the magnetic field, dominant  
within few stellar radii from the star, where the maser emission should be generated, 
inhibits the onset of the cyclotron maser instability in $\sigma$\,Ori\,E.
\end{abstract}


\section{Introduction}
Magnetic chemically peculiar (MCP) stars present photometric, spectroscopic
and magnetic variability with a single period. As to the magnetic field, the
most important observable is the effective or longitudinal magnetic field ($B_{\rm e}$),
that is the average over the stellar visible disk of the longitudinal components of the magnetic field \citep*{Leone00}.
To explain the previous phenomenology
\citet{babcock49} and \citet{stibbs50} proposed the presence of a magnetic
dipole, whose axis is tilted with respect to the rotational axis, the so called Oblique Rotator Model,
and a non-uniform distribution of some chemical elements on the stellar
surface which rotates rigidly.
Spectral, light and magnetic variability
would be a consequence  of stellar rotation. Later, it has been cleared
that the MCP stars can present multipolar magnetic fields. \citet*{bychkov_etal05}
collect all available 
MCP magnetic curves and they show that
in many cases their shape is  more complex then a simple sinusoidal wave.

Non-thermal radio emission is observed from about 25\% of the MCP stars \citep*{leone_etal94}. 
In accord with the oblique rotator model, the radio emission is also variable as a consequence
of the stellar rotation \citep{leone91}, suggesting that the radio emission 
arises from a stable optically thick co-rotating magnetosphere. 
Radio emission is ascribed to a radiatively-driven stellar wind. The gas flow 
brakes the magnetic field lines far from the star (Alfv\'en surface)
forming current sheets, where electrons are accelerated to the mildly relativistic regime. Energetic electrons propagate back, along 
thin magnetospheric layers named middle magnetosphere, 
to the inner magnetospheric regions radiating 
by gyrosynchrotron emission mechanism at the radio wavelengths
\citep{trigilio_etal04,leto_etal06}.

Out of MCP stars, CU\,Virginis (HD\,124224 = HR\,5313)
is the only known source characterized by 
broadband, highly polarized and time-stable pulses at 1.4 and 2.5~GHz
\citep{trigilio_etal00,trigilio_etal08,trigilio_etal11,ravi_etal10,lo_etal12}.
\citet{stevens_george10}
have also reported a pulse detection at 610 MHz.
Radio pulses stand out for 1 dex over the continuous emission,
{ the single pulse duration ranges from 5 to 10\% of the rotational period}
and are observed in coincidence with the null values of $B_{\rm e}$, that is
when the magnetic dipole axis is perpendicular to the line of sight.

Such a behavior has been ascribed to the electron cyclotron maser
(ECM) emission mechanism 
powered by an anisotropic pitch-angle
(angle between electron velocity and local magnetic field) distribution that the
non-thermal electrons, responsible of the gyrosynchrotron stellar radio emission, 
can develop propagating in a magnetic flux tube
towards regions of increasing magnetic field strength. 
Electrons with an initial large pitch-angle are soon reflected outward due to the magnetic mirroring, whereas
electrons with a small pitch angle can reach the inner magnetospheric layers
where they are thermalized in the dense plasma. The reflected electron population 
is characterized by 
a pitch angle distribution deprived by the electrons with a small pitch angle.
This mechanism 
amplifies the extraordinary magneto-ionic mode producing nearly 100\%
circularly polarized radiation at frequencies very close to the first or
second harmonic of the local gyro-frequency 
($\nu_{\mathrm B}=2.8 \times 10^{-3} B/{\m G}$ GHz),
in a direction almost perpendicular to the 
local magnetic field lines \citep{melrose_dulk82}. 
However the fundamental harmonic is probably suppressed by
the gyromagnetic absorption. 

The ECM mechanism has been considered to account for 
the strongly polarized, intense and narrow band short time spikes
observed in the Sun \citep{willson85,winglee_dulk86},
dMe flare stars \citep{lang_etal83,lang_willson88,abada-simon_etal94,abada-simon_etal97} 
and RS\,CVn binaries 
\citep*{slee_etal84,osten_etal04}. ECM explains 
the planetary low frequency emission, in particular 
the Jupiter decametric radiation
and the Earth's Auroral
Kilometric Radiation (AKR)  \citep{treumann06}.

The fully polarized pulses observed on CU\,Vir are broadband
and persistent over long timescale (years).
The large bandwidth is a consequence of the wide range of the magnetic field strength
in the region where the maser amplification occurs.
The observed maser emission is the superimposition of narrowband emission from different 
rings above the magnetic poles.
The continuos supply of non-thermal electrons, developing a loss-cone anisotropy,
maintains stable the electron cyclotron maser emission.
Following the tangent plane beaming model, proposed for  the AKR
\citep*{mutel08} and successfully applied to explain the narrow peaks observed on CU\,Vir 
\citep{trigilio_etal11},
the amplified radiation is beamed tangentially 
to the polar ring where the cyclotron maser instability takes place.
{ Then, during the propagation through the denser magnetized plasma of the inner magnetosphere,
the radiation is refracted upward by a few degrees \citep{trigilio_etal11,lo_etal12}.
Since the magnitude of this angle depends on frequency, the radiation is detected in different
moments during the rotation of the star, causing a frequency drift of the observed pulses.}

Highly polarized broad-band radio pulses, still explained as ECM, have
been also recognized in the ultra cool main sequence dwarf stars
\citep{berger_etal01,burgasser_putman05,hallinan_etal06,hallinan_etal07,hallinan_etal08}.
Despite the great difference of the physical characteristics between these two extreme
classes of stars, late M and MCP stars show similar behavior at the radio 
wavelengths. The existence of a well ordered and stable axisymmetric magnetic field in the
rapidly rotating fully convective M type stars
\citep{donati_etal06a,donati_etal08, morin_etal08a,morin_etal08b} can be the reason of  the observed similarity.

The study of the radio pulses from CU\,Vir also provided
the possibility to evidence variations of the stellar rotational period with a high degree of confidence 
\citep{trigilio_etal08,trigilio_etal11}. 
Discovering the CU\,Virginis type coherent emission in other MCP stars would provide an useful tool for the study of the angular momentum evolution in this kind of stars.

On the basis of the results gathered from CU\,Vir, the possibility to
observe the same type of coherent emission from other MCP stars 
requires an appropriate stellar magnetic field geometry, with
$B_{\rm e}$ presenting  at least a null value during the stellar rotation. 
Among the already known MCP stars presenting radio emission 
$\sigma$\,Ori\,E (HD\,37479 = HR\,1932) is an ideal candidate
to search  for  the presence of this type of coherent emission. 
Its magnetic field presents the appropriate geometry, being the 
rotation axis inclination $i=72^{\circ}$ and magnetic axis obliquity 
$\beta=56^{\circ}$ \citep{bohlender_etal87}. 

The coherent pulses have been observed from CU\,Vir
($B_{\m p} = 3000$ G, \citet{trigilio_etal00}) at low frequencies
($<$ 2.5 GHz).
We expect that ECM emission from $\sigma$\,Ori\,E
could be observed at higher frequencies, as a consequence of the stronger magnetic field
($B_{\m p} = 6800$ G, \citet{trigilio_etal04}).
In this paper we present new multifrequency (1.4, 5, 8.4, 15, 22 and 43 GHz) 
VLA observations of $\sigma$\,Ori\,E. In addition,
to obtain as complete as possible radio light-curves, we have retrieved all unpublished
VLA archive data.
The primary aim is to search for amplified emission strongly polarized, 
but we can also probe different layers of the $\sigma$\,Ori\,E magnetosphere,
because of the dependence of the frequency of the gyrosynchrotron 
radio emission from the magnetic field strength.

\section{$\sigma$\,Ori\,E}
{ $\sigma$\,Ori\,E is a magnetic helium-strong star of spectral type B2V,  mass $M_{\ast}=8.9$ M$_{\odot}$ \citep*{hunger_etal89},
radius $R_{\ast}=4.2$ R$_{\odot}$ \citep{shore_brown90}, located at a distance of about 350 pc \citep{hipparcos}.
It is significantly hotter then CU\,Vir (spectral type A0V), 
this involves a stronger radiatively driven wind and, therefore, a greater mass loss rate. 
On the other hand, the polar magnetic field strength of $\sigma$\,Ori\,E is 6800 G, versus 3000 G of CU\,Vir, 
and the rotational period is 1.19$^\mathrm{d}$, versus 0.52$^\mathrm{d}$.
All these two factors influence the confinement of the wind, resulting that the two stars
have similar magnetospheres, both with Alfv\'en radius about 15 R$_{\ast}$ \citep{trigilio_etal04,leto_etal06}.
This reinforces our decision to search evidence of CU\,Vir-type coherent emission from $\sigma$\,Ori\,E.}

The long history of the photometric, spectroscopic and magnetic variability presented by $\sigma$\,Ori\,E
is documented in the {\it Catalog of observed periods for the AP stars}
by \citet{Catalano84} and its supplements \citep*{Catalano88,Catalano91a,Catalano93}.  
\citet{townsend_etal10} have 
concluded that $\sigma$\,Ori\,E
is spinning down because of the magnetic braking and supplied the ephemeris:
\\

\noindent
$\mathrm {HJD}=2\,442\,778.829+
        1.1908229 E + 1.44 \times 10^{-9} E ^2 ~~\mathrm{[days]}
$\\

\noindent
{ referred to the deeper light minimum. The spin down of $\sigma$\,Ori\,E was already theoretically predicted
by magnetohydrodynamics simulations \citep*{uddoula_etal09}.
}

\begin{figure}
\resizebox{\hsize}{!}{\includegraphics{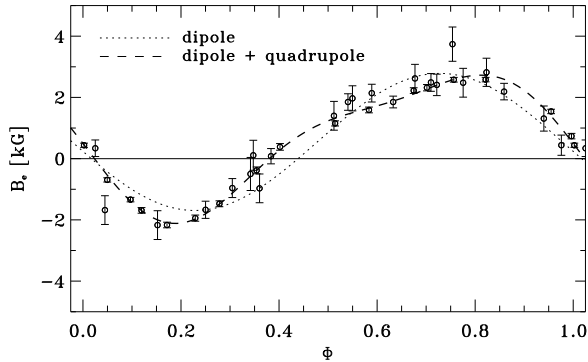}}
\caption{Fit of the literature magnetic field measurements -- open circles.
Legend of line style: single wave fit -- dot line,
double wave fit -- dashed line. The solid horizontal line coincides with the null effective surface magnetic field.}
\label{fit_b}
\end{figure}

The global magnetic field of $\sigma$\,Ori\,E is steady over three decades \citep{oksala_etal10}
and since \citet{landstreet_borra78} a pure magnetic dipole is assumed.
 Fig.~\ref{fit_b} shows the magnetic field measurements of $\sigma$\,Ori\,E by
\citet{landstreet_borra78}, \citet{bohlender_etal87} and \citet{oksala_etal10} phased with the previous
ephemeris. A data fit with a single wave equation:
\begin{displaymath}
B_{\mathrm e}=B_0 + B_1 \sin 2\pi (\Phi -\Phi_0)
\end{displaymath}

\noindent
 gives a reduced $\chi ^2 = 12.4$. Here $B_0=0.55 \pm 0.02$ [kG],
$B_1=2.24 \pm 0.03$ [kG] and $\Phi_0=0.478 \pm 0.002$.

From long time it is known that the variability of MCP stars
can be accurately modeled through a sinusoidal wave and its first harmonic
\citep*{Catalano91b}. Indeed a fit of magnetic data with the equation:
\begin{displaymath}
B_{\mathrm e}=B_0 + B_1 \sin 2\pi (\Phi -\Phi_0) + B_2 \sin 2\pi (2\Phi - \Phi_1)
\end{displaymath}

\noindent
results in a reduced $\chi ^2 = 1.6$, suggesting the 
non-dipolar nature of the $\sigma$\,Ori\,E magnetic field.
Here $B_0=0.62 \pm 0.02$ [kG], $B_1=2.21 \pm 0.03$ [kG], 
$B_2=0.63 \pm 0.03$ [kG], $\Phi_0=0.473 \pm 0.002$ and $\Phi_1=0.563 \pm 0.007$. 

\cite*{Landolfi_etal98} shown that the $B_{\mathrm e}$ variation due to a pure quadrupole is about 10\% of the variability
due to a dipole of equal polar strength. This means that $\sigma$\,Ori\,E would present a quadrupole component
at least two times stronger than the dipole one.

The presence of multipolar magnetic components in $\sigma$\,Ori\,E
is testified by the asymmetric emission features
on the H${\alpha}$  wings. These  periodically variable features have been ascribed to two
circumstellar plasma clouds 
trapped in the magnetosphere and 
co-rotating with the star \citep{landstreet_borra78};
trapped material was also recognized from UV observations \citep{smith_groote01}.
To explain why one of the two feature is stronger than the other, \citet{Groote_hunger82} suggested an asymmetry in the two clouds. In the case of $\delta$\,Ori\,C, an MCP star twin of $\sigma$\,Ori\,E, also presenting similar asymmetric emission features on the H${\alpha}$  wings, 
\citet{leone_etal10} mapped the circumstellar matter distribution and ascribed the asymmetry between the two clouds to the magnetic quadrupole component necessary to explain the Stokes V profiles of the spectral lines.

We conclude that the $\sigma$\,Ori\,E phenomenology cannot be modeled as a simple dipole and that higher order components are
necessary to explain the observational data.

\begin{table} 
\caption{VLA observing log.} 
\label{VLA_log} 
\begin{center} 
\begin{tabular}{cccccc} 
\hline           
\hline 

\s Code   &\s Frequencies    &\s Epoch    &\s conf.   &\s Flux cal   &\s 
Phase cal  \\ 
\s                       &\s [GHz]              &   &   &   &  \\ 
\hline 
\s AL346            &\s 5/15/43         &\s 95-Apr                     &\s D   
&\s 3C48   &\s 0541$-$056  \\ 
\s AT233            &\s 1.4/5           &\s 99-Oct   &\s AB   &\s 3C48     
&\s 0541$-$056  \\ 
\s AL568            &\s 8.4/15         &\s 02-May                    &\s AB 
 &\s 3C286   &\s 0541$-$056  \\ 
\s AL618            &\s 5/15/22/43         &\s 04-Jan                   
&\s BC   &\s 3C286   &\s 0541$-$056  \\ 
\hline 
\multicolumn{6}{l}{\s Archive VLA data}\\ 
\hline 
\s AL267            &\s 5/8.4/15         &\s 92-Oct                     &\s 
A   &\s 3C286   &\s 0541$-$056  \\ 
\s AL348            &\s 5/8.4/15/22/43         &\s 95-Mar                     &\s 
D   &\s 3C286   &\s 0532+075  \\ 
\s AL372            &\s 5/8.4/15         &\s 96-Mar                     &\s 
C   &\s 3C48   &\s 0541$-$056  \\

\hline             
\end{tabular}     
\end{center} 

\end{table} 

\begin{figure}
\resizebox{\hsize}{!}{\includegraphics{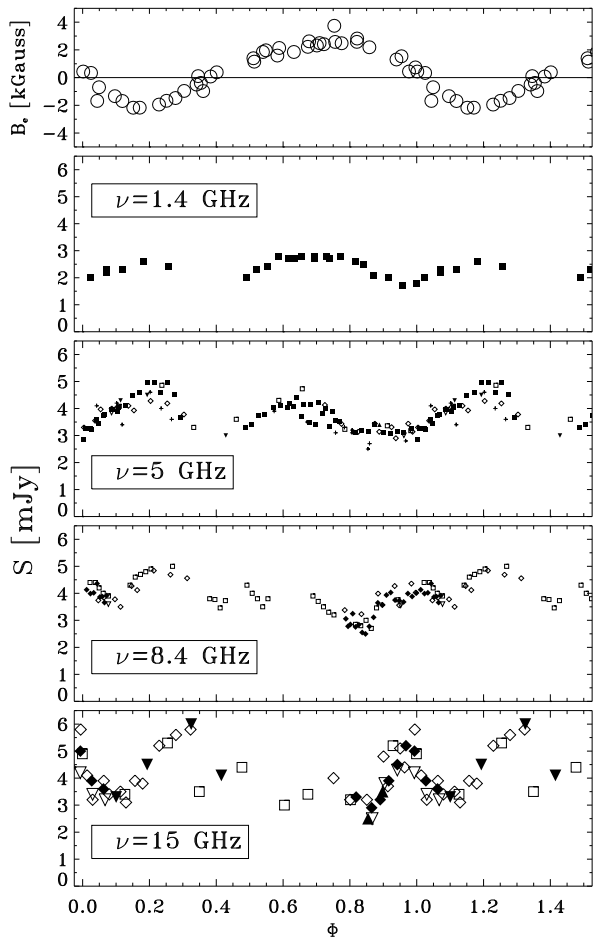}}
\caption{VLA measurements of $\sigma$\,Ori\,E plotted versus the rotation phase $\Phi$.
Legend of data: AL267 ($\square$), AL348 ($\triangledown$), AL346 ($\blacktriangledown$), AL372 ($\Diamond$), AT233 ($\blacksquare$), AL568 ($\blacklozenge$) and AL618 ($\blacktriangle$).
Literature data at 5 GHz are indicated as ($+$).
Symbol size is equal to the flux error.
Top panel reports the literature magnetic field curve.
}
\label{dati}
\end{figure}

\begin{table*}
\caption{New and archival VLA observations.}
\label{tab}
\begin{center}
\begin{tabular}{@{}c @{~} c @{~} c | c @{~} c @{~} c | c @{~} c @{~} c | c @{~} c @{~} c | c @{~} c @{~} c@{}}
\hline          
\hline

 \s JD       &\s S$_{\m I}$\ty(S$_{\m V}$)    &\s $\sigma$     &\s JD       &\s S$_{\m I}$\ty(S$_{\m V}$)  &\s $\sigma$    &\s JD       &\s S$_{\m I}$\ty(S$_{\m V}$)    &\s $\sigma$    &\s JD       &\s S$_{\m I}$\ty(S$_{\m V}$)     &\s $\sigma$   &\s JD       &\s S$_{\m I}$\ty(S$_{\m V}$)     &\s $\sigma$  \\

    \ty2400000+         &\ty mJy &\ty mJy              &\ty 2400000+           &\ty mJy              &\ty mJy   &\ty 2400000+           &\ty mJy              &\ty mJy   &\ty 2400000+           &\ty mJy              &\ty mJy    &\ty 2400000+           &\ty mJy              &\ty mJy    \\
\hline
       \cline{1-3}
 \multicolumn{3}{c}{\s $\nu$=1.4 GHz}                 &\s 50171.677$^{\m d}$ &\s 4.10\ty($-0.03$) &\ty 0.07  &\s 51461.958$^{\m e}$ &\s 4.42\ty($+0.13$) &\ty 0.05   &\s 50171.629$^{\m d}$ &\s 3.78\ty($-0.27$) &\ty 0.05   &\s 49827.591$^{\m c}$ &\s 4.50\ty($+0.10$) &\ty 0.40 \\       
      \cline{1-3}
\s 51453.833$^{\m e}$ &\s 2.60\ty($+0.10$) &\ty 0.15   &\s 50172.377$^{\m d}$ &\s 4.10\ty($+0.30$) &\ty 0.10  &\s 51461.982$^{\m e}$ &\s 4.14\ty($+0.20$) &\ty 0.06   &\s 50171.688$^{\m d}$ &\s 4.27\ty($-0.14$) &\ty 0.06  &\s 49833.434$^{\m c}$ &\s 3.30\ty($-0.40$) &\ty 0.30 \\           
\s 51453.862$^{\m e}$ &\s 2.50\ty($+0.10$) &\ty 0.15   &\s 50172.437$^{\m d}$ &\s 3.43\ty($+0.22$) &\ty 0.08  &\s 51462.009$^{\m e}$ &\s 4.14\ty($+0.15$) &\ty 0.06   &\s 50172.448$^{\m d}$ &\s 3.38\ty($+0.43$) &\ty 0.06  &\s 49837.382$^{\m c}$ &\s 4.10\ty($+0.10$) &\ty 0.30 \\            
\s 51453.896$^{\m e}$ &\s 2.10\ty($+0.00$) &\ty 0.15   &\s 50172.496$^{\m d}$ &\s 3.14\ty($+0.05$) &\ty 0.08  &\s 51462.037$^{\m e}$ &\s 4.21\ty($+0.35$) &\ty 0.06   &\s 50172.507$^{\m d}$ &\s 3.23\ty($+0.29$) &\ty 0.06  &\s 50170.358$^{\m d}$ &\s 3.20\ty($-0.30$) &\ty 0.10 \\             
\s 51453.951$^{\m e}$ &\s 2.00\ty($+0.00$) &\ty 0.20   &\s 50172.555$^{\m d}$ &\s 3.44\ty($+0.04$) &\ty 0.08  &\s 51462.061$^{\m e}$ &\s 4.00\ty($+0.24$) &\ty 0.05   &\s 50172.567$^{\m d}$ &\s 3.99\ty($+0.10$) &\ty 0.05  &\s 50170.418$^{\m d}$ &\s 3.40\ty($-0.30$) &\ty 0.10 \\             
\s 51453.998$^{\m e}$ &\s 1.70\ty($+0.10$) &\ty 0.20   &\s 50172.615$^{\m d}$ &\s 3.31\ty($-0.01$) &\ty 0.07  &\s 51462.085$^{\m e}$ &\s 3.88\ty($+0.15$) &\ty 0.06   &\s 50172.626$^{\m d}$ &\s 4.27\ty($+0.05$) &\ty 0.05  &\s 50170.477$^{\m d}$ &\s 3.10\ty($-0.30$) &\ty 0.10 \\             
\s 51454.050$^{\m e}$ &\s 1.80\ty($-0.10$) &\ty 0.20   &\s 50172.674$^{\m d}$ &\s 3.44\ty($+0.08$) &\ty 0.07  &\s 51462.111$^{\m e}$ &\s 3.53\ty($+0.27$) &\ty 0.05   &\s 50172.685$^{\m d}$ &\s 4.36\ty($+0.04$) &\ty 0.06  &\s 50170.537$^{\m d}$ &\s 3.80\ty($-0.40$) &\ty 0.10 \\            
\s 51454.135$^{\m e}$ &\s 2.30\ty($+0.00$) &\ty 0.20   &\s 51453.831$^{\m e}$ &\s 3.10\ty($+0.10$) &\ty 0.10  &\s 53031.789$^{\m g}$ &\s 3.40\ty($+0.05$) &\ty 0.08   &\s 52417.192$^{\m f}$ &\s 3.10\ty($+0.50$) &\ty 0.20  &\s 50170.596$^{\m d}$ &\s 5.20\ty($-0.40$) &\ty 0.10 \\  
                                                                                                                         \cline{7-9}         
\s 51455.827$^{\m e}$ &\s 2.00\ty($-0.10$) &\ty 0.15   &\s 51453.855$^{\m e}$ &\s 3.20\ty($+0.20$) &\ty 0.20  &\multicolumn{3}{c}{\s $\nu$=8.4 GHz}                   &\s 52417.200$^{\m f}$ &\s 2.80\ty($+0.70$) &\ty 0.20  &\s 50170.655$^{\m d}$ &\s 5.60\ty($-0.20$) &\ty 0.10 \\ 
                                                                                                                         \cline{7-9}         
\s 51455.862$^{\m e}$ &\s 2.30\ty($+0.10$) &\ty 0.10   &\s 51453.878$^{\m e}$ &\s 3.10\ty($+0.10$) &\ty 0.10  &\s 48906.869$^{\m a}$ &\s 4.40\ty($+0.10$) &\ty 0.10   &\s 52417.208$^{\m f}$ &\s 2.80\ty($+0.30$) &\ty 0.20  &\s 50170.707$^{\m d}$ &\s 5.80\ty($-0.20$) &\ty 0.20 \\            
\s 51455.902$^{\m e}$ &\s 2.40\ty($+0.00$) &\ty 0.10   &\s 51453.902$^{\m e}$ &\s 3.40\ty($+0.10$) &\ty 0.10  &\s 48906.887$^{\m a}$ &\s 4.40\ty($+0.00$) &\ty 0.10   &\s 52417.217$^{\m f}$ &\s 3.20\ty($+0.30$) &\ty 0.20  &\s 50171.411$^{\m d}$ &\s 3.70\ty($+0.20$) &\ty 0.10 \\            
\s 51455.941$^{\m e}$ &\s 2.80\ty($-0.10$) &\ty 0.10   &\s 51453.934$^{\m e}$ &\s 3.11\ty($-0.02$) &\ty 0.08  &\s 48906.901$^{\m a}$ &\s 4.20\ty($-0.30$) &\ty 0.10   &\s 52417.227$^{\m f}$ &\s 2.80\ty($+0.60$) &\ty 0.20  &\s 50171.470$^{\m d}$ &\s 4.40\ty($+0.00$) &\ty 0.10 \\            
\s 51455.998$^{\m e}$ &\s 2.70\ty($+0.10$) &\ty 0.10   &\s 51453.957$^{\m e}$ &\s 3.09\ty($+0.11$) &\ty 0.07  &\s 48906.915$^{\m a}$ &\s 4.00\ty($-0.10$) &\ty 0.10   &\s 52417.238$^{\m f}$ &\s 2.80\ty($+0.30$) &\ty 0.20  &\s 50171.530$^{\m d}$ &\s 4.10\ty($+0.00$) &\ty 0.10 \\            
\s 51456.067$^{\m e}$ &\s 2.70\ty($+0.10$) &\ty 0.10   &\s 51453.981$^{\m e}$ &\s 3.16\ty($+0.13$) &\ty 0.07  &\s 48906.934$^{\m a}$ &\s 3.90\ty($-0.20$) &\ty 0.10   &\s 52417.250$^{\m f}$ &\s 2.50\ty($+0.20$) &\ty 0.20  &\s 50171.589$^{\m d}$ &\s 3.90\ty($-0.30$) &\ty 0.10 \\            
\s 51456.120$^{\m e}$ &\s 2.70\ty($+0.00$) &\ty 0.10   &\s 51454.004$^{\m e}$ &\s 3.12\ty($+0.01$) &\ty 0.06  &\s 48907.011$^{\m a}$ &\s 4.30\ty($-0.20$) &\ty 0.10   &\s 52417.262$^{\m f}$ &\s 2.50\ty($+0.30$) &\ty 0.10  &\s 50171.648$^{\m d}$ &\s 3.50\ty($-0.40$) &\ty 0.10 \\            
\s 51456.162$^{\m e}$ &\s 2.80\ty($+0.00$) &\ty 0.10   &\s 51454.032$^{\m e}$ &\s 3.25\ty($-0.07$) &\ty 0.05  &\s 48907.030$^{\m a}$ &\s 4.60\ty($-0.10$) &\ty 0.10   &\s 52417.276$^{\m f}$ &\s 2.80\ty($+0.20$) &\ty 0.10  &\s 50171.700$^{\m d}$ &\s 3.90\ty($-0.80$) &\ty 0.20 \\            
\s 51458.841$^{\m e}$ &\s 2.00\ty($-0.10$) &\ty 0.08   &\s 51454.060$^{\m e}$ &\s 3.28\ty($-0.05$) &\ty 0.06  &\s 48907.048$^{\m a}$ &\s 4.70\ty($-0.10$) &\ty 0.10   &\s 52417.291$^{\m f}$ &\s 3.10\ty($+0.10$) &\ty 0.10  &\s 50172.408$^{\m d}$ &\s 4.00\ty($+0.80$) &\ty 0.20 \\            
\s 51458.899$^{\m e}$ &\s 2.20\ty($+0.00$) &\ty 0.10   &\s 51454.083$^{\m e}$ &\s 3.22\ty($+0.01$) &\ty 0.06  &\s 48907.067$^{\m a}$ &\s 4.80\ty($+0.00$) &\ty 0.10   &\s 52417.306$^{\m f}$ &\s 3.60\ty($+0.00$) &\ty 0.10  &\s 50172.467$^{\m d}$ &\s 3.20\ty($+0.20$) &\ty 0.20 \\            
\s 51458.956$^{\m e}$ &\s 2.30\ty($-0.06$) &\ty 0.07   &\s 51454.107$^{\m e}$ &\s 3.44\ty($-0.05$) &\ty 0.06  &\s 48907.086$^{\m a}$ &\s 4.90\ty($+0.20$) &\ty 0.10   &\s 52417.321$^{\m f}$ &\s 3.60\ty($-0.20$) &\ty 0.10  &\s 50172.527$^{\m d}$ &\s 3.20\ty($+0.20$) &\ty 0.10 \\            
\s 51459.031$^{\m e}$ &\s 2.60\ty($+0.04$) &\ty 0.08   &\s 51454.130$^{\m e}$ &\s 3.78\ty($+0.00$) &\ty 0.06  &\s 48907.163$^{\m a}$ &\s 5.00\ty($+0.10$) &\ty 0.10   &\s 52417.337$^{\m f}$ &\s 3.90\ty($+0.10$) &\ty 0.10  &\s 50172.586$^{\m d}$ &\s 4.80\ty($-0.10$) &\ty 0.10 \\            
\s 51459.122$^{\m e}$ &\s 2.40\ty($+0.02$) &\ty 0.08   &\s 51454.154$^{\m e}$ &\s 3.96\ty($-0.04$) &\ty 0.07  &\s 48908.854$^{\m a}$ &\s 3.90\ty($+0.20$) &\ty 0.10   &\s 52417.352$^{\m f}$ &\s 4.00\ty($+0.00$) &\ty 0.10  &\s 50172.645$^{\m d}$ &\s 5.10\ty($+0.10$) &\ty 0.10 \\            
\s 51461.929$^{\m e}$ &\s 2.70\ty($+0.00$) &\ty 0.10   &\s 51454.176$^{\m e}$ &\s 3.87\ty($-0.13$) &\ty 0.08  &\s 48908.873$^{\m a}$ &\s 3.70\ty($+0.10$) &\ty 0.10   &\s 52417.367$^{\m f}$ &\s 3.70\ty($-0.10$) &\ty 0.10  &\s 50172.697$^{\m d}$ &\s 5.80\ty($-0.10$) &\ty 0.20 \\            
\s 51461.976$^{\m e}$ &\s 2.80\ty($+0.20$) &\ty 0.10   &\s 51455.821$^{\m e}$ &\s 3.30\ty($+0.20$) &\ty 0.10  &\s 48908.891$^{\m a}$ &\s 3.50\ty($+0.40$) &\ty 0.10   &\s 52417.383$^{\m f}$ &\s 3.73\ty($-0.06$) &\ty 0.08  &\s 52417.229$^{\m f}$ &\s 3.30\ty($+0.70$) &\ty 0.15 \\            
\s 51462.021$^{\m e}$ &\s 2.80\ty($+0.10$) &\ty 0.10   &\s 51455.845$^{\m e}$ &\s 3.41\ty($+0.15$) &\ty 0.08  &\s 48908.910$^{\m a}$ &\s 3.30\ty($+0.20$) &\ty 0.10   &\s 52417.398$^{\m f}$ &\s 3.70\ty($+0.10$) &\ty 0.10  &\s 52417.284$^{\m f}$ &\s 2.90\ty($+0.40$) &\ty 0.20 \\           
\s 51462.067$^{\m e}$ &\s 2.80\ty($+0.00$) &\ty 0.10   &\s 51455.868$^{\m e}$ &\s 3.75\ty($+0.01$) &\ty 0.08  &\s 48908.929$^{\m a}$ &\s 3.20\ty($+0.40$) &\ty 0.10   &\s 52417.413$^{\m f}$ &\s 4.00\ty($-0.10$) &\ty 0.10  &\s 52417.315$^{\m f}$ &\s 3.20\ty($+0.00$) &\ty 0.15 \\          
      \cline{1-3}
 \multicolumn{3}{c}{\s $\nu$=5 GHz}                    &\s 51455.892$^{\m e}$ &\s 3.80\ty($+0.20$) &\ty  0.10 &\s 48909.006$^{\m a}$ &\s 2.80\ty($+0.20$) &\ty 0.10   &\s 52417.428$^{\m f}$ &\s 3.90\ty($-0.10$) &\ty 0.10  &\s 52417.345$^{\m f}$ &\s 3.90\ty($+0.10$) &\ty 0.15 \\          
      \cline{1-3}
\s 48906.821$^{\m a}$ &\s 3.20\ty($+0.00$) &\ty 0.10   &\s 51455.923$^{\m e}$ &\s 4.03\ty($+0.18$) &\ty 0.08  &\s 48909.024$^{\m a}$ &\s 2.80\ty($+0.20$) &\ty 0.10   &\s 52417.444$^{\m f}$ &\s 4.00\ty($+0.00$) &\ty 0.10  &\s 52417.376$^{\m f}$ &\s 4.50\ty($-0.40$) &\ty 0.15 \\           
\s 48906.973$^{\m a}$ &\s 4.10\ty($+0.20$) &\ty 0.10   &\s 51455.947$^{\m e}$ &\s 4.13\ty($+0.14$) &\ty 0.08  &\s 48909.043$^{\m a}$ &\s 3.00\ty($+0.10$) &\ty 0.10   &\s 52417.459$^{\m f}$ &\s 4.10\ty($-0.10$) &\ty 0.10  &\s 52417.406$^{\m f}$ &\s 5.20\ty($+0.50$) &\ty 0.15 \\            
\s 48907.125$^{\m a}$ &\s 4.90\ty($-0.10$) &\ty 0.10   &\s 51455.971$^{\m e}$ &\s 4.04\ty($+0.18$) &\ty 0.07  &\s 48909.062$^{\m a}$ &\s 2.70\ty($+0.20$) &\ty 0.10   &\s 52417.473$^{\m f}$ &\s 4.00\ty($-0.10$) &\ty 0.10  &\s 52417.437$^{\m f}$ &\s 5.00\ty($+0.00$) &\ty 0.15 \\            
\s 48908.816$^{\m a}$ &\s 4.70\ty($+0.20$) &\ty 0.10   &\s 51455.994$^{\m e}$ &\s 4.09\ty($+0.15$) &\ty 0.08  &\s 48909.081$^{\m a}$ &\s 3.50\ty($+0.00$) &\ty 0.10   &\s 52417.484$^{\m f}$ &\s 4.00\ty($+0.00$) &\ty 0.10  &\s 52417.477$^{\m f}$ &\s 3.90\ty($-0.10$) &\ty 0.20 \\            
\s 48908.968$^{\m a}$ &\s 3.20\ty($+0.00$) &\ty 0.10   &\s 51456.022$^{\m e}$ &\s 3.71\ty($+0.17$) &\ty 0.08  &\s 48909.157$^{\m a}$ &\s 3.80\ty($+0.00$) &\ty 0.10   &\s 52417.496$^{\m f}$ &\s 4.40\ty($-0.10$) &\ty 0.10  &\s 52417.519$^{\m f}$ &\s 3.60\ty($-0.30$) &\ty 0.20 \\           
\s 48909.119$^{\m a}$ &\s 3.40\ty($+0.10$) &\ty 0.10   &\s 51456.050$^{\m e}$ &\s 3.49\ty($+0.26$) &\ty 0.07  &\s 48910.867$^{\m a}$ &\s 3.80\ty($+0.00$) &\ty 0.10   &\s 52417.506$^{\m f}$ &\s 3.90\ty($-0.20$) &\ty 0.20  &\s 53031.748$^{\m f}$ &\s 2.50\ty($+0.30$) &\ty 0.20 \\           
\s 48910.811$^{\m a}$ &\s 3.30\ty($+0.00$) &\ty 0.15   &\s 51456.073$^{\m e}$ &\s 3.42\ty($+0.33$) &\ty 0.07  &\s 48910.886$^{\m a}$ &\s 3.77\ty($-0.05$) &\ty 0.08   &\s 52417.515$^{\m f}$ &\s 3.90\ty($+0.00$) &\ty 0.20  &\s 53031.800$^{\m f}$ &\s 3.50\ty($+0.50$) &\ty 0.20 \\           
                                                                                                                                                                                                                                              \cline{13-15}       
\s 48910.962$^{\m a}$ &\s 3.60\ty($+0.10$) &\ty 0.15   &\s 51456.097$^{\m e}$ &\s 3.81\ty($+0.20$) &\ty 0.08  &\s 48910.905$^{\m a}$ &\s 3.46\ty($+0.03$) &\ty 0.08   &\s 52417.523$^{\m f}$ &\s 3.70\ty($+0.00$) &\ty 0.20   &\multicolumn{3}{c}{\s $\nu$=22 GHz }                \\           
                                                                                                                                                                                                                                              \cline{13-15}       
\s 48911.114$^{\m a}$ &\s 4.30\ty($+0.20$) &\ty 0.15   &\s 51456.120$^{\m e}$ &\s 3.32\ty($+0.06$) &\ty 0.08  &\s 48910.924$^{\m a}$ &\s 3.73\ty($-0.11$) &\ty 0.08   &\s 52417.531$^{\m f}$ &\s 3.90\ty($+0.10$) &\ty 0.20  &\s 49803.407$^{\m b}$ &\s 2.70\ty($+0.10$) &\ty 0.30 \\           
                                                                                                                                                                               \cline{10-12}       
\s 49803.498$^{\m b}$ &\s 3.00\ty($-0.20$) &\ty 0.20   &\s 51456.144$^{\m e}$ &\s 3.55\ty($+0.17$) &\ty 0.08  &\s 48911.000$^{\m a}$ &\s 4.30\ty($+0.10$) &\ty 0.10   &\multicolumn{3}{c}{\s $\nu$=15 GHz}                   &\s 49803.452$^{\m b}$ &\s 3.50\ty($-0.60$) &\ty 0.30 \\           
                                                                                                                                                                               \cline{10-12}       
\s 49803.647$^{\m b}$ &\s 3.80\ty($-0.40$) &\ty 0.20   &\s 51458.818$^{\m e}$ &\s 2.84\ty($-0.01$) &\ty 0.08  &\s 48911.019$^{\m a}$ &\s 4.00\ty($+0.10$) &\ty 0.10   &\s 48906.841$^{\m a}$ &\s 4.90\ty($-0.10$) &\ty 0.30  &\s 49803.513$^{\m b}$ &\s 4.20\ty($+0.50$) &\ty 0.30 \\           
\s 49827.591$^{\m c}$ &\s 4.50\ty($+0.10$) &\ty 0.10   &\s 51458.841$^{\m e}$ &\s 3.24\ty($+0.06$) &\ty 0.07  &\s 48911.038$^{\m a}$ &\s 3.80\ty($+0.00$) &\ty 0.10   &\s 48906.993$^{\m a}$ &\s 3.40\ty($-0.70$) &\ty 0.40  &\s 49803.557$^{\m b}$ &\s 3.30\ty($+0.50$) &\ty 0.20 \\           
\s 49833.450$^{\m c}$ &\s 4.30\ty($-0.10$) &\ty 0.10   &\s 51458.865$^{\m e}$ &\s 3.59\ty($+0.06$) &\ty 0.07  &\s 48911.057$^{\m a}$ &\s 3.50\ty($+0.30$) &\ty 0.10   &\s 48907.145$^{\m a}$ &\s 5.30\ty($-0.40$) &\ty 0.40  &\s 49803.602$^{\m b}$ &\s 2.90\ty($-0.60$) &\ty 0.30 \\           
\s 49837.398$^{\m c}$ &\s 3.00\ty($+0.00$) &\ty 0.10   &\s 51458.889$^{\m e}$ &\s 3.73\ty($-0.07$) &\ty 0.07  &\s 48911.075$^{\m a}$ &\s 3.80\ty($+0.10$) &\ty 0.10   &\s 48908.836$^{\m a}$ &\s 3.40\ty($+0.90$) &\ty 0.40  &\s 53031.777$^{\m g}$ &\s 3.20\ty($+0.60$) &\ty 0.10 \\           
\s 50170.328$^{\m d}$ &\s 3.30\ty($+0.00$) &\ty 0.10   &\s 51458.920$^{\m e}$ &\s 3.96\ty($-0.10$) &\ty 0.08  &\s 49803.487$^{\m b}$ &\s 3.55\ty($+0.00$) &\ty 0.10   &\s 48908.988$^{\m a}$ &\s 3.20\ty($+0.90$) &\ty 0.40  &\s 53031.828$^{\m g}$ &\s 4.60\ty($+0.00$) &\ty 0.10 \\           
\s 50170.387$^{\m d}$ &\s 3.97\ty($-0.04$) &\ty 0.08   &\s 51458.944$^{\m e}$ &\s 4.05\ty($-0.06$) &\ty 0.07  &\s 49803.636$^{\m b}$ &\s 3.60\ty($-0.20$) &\ty 0.10   &\s 48909.139$^{\m a}$ &\s 5.20\ty($+0.50$) &\ty 0.40  &\s 53031.859$^{\m g}$ &\s 5.30\ty($+0.20$) &\ty 0.10 \\           
                                                                                                                                                                                                                                              \cline{13-15}       
\s 50170.446$^{\m d}$ &\s 4.05\ty($-0.10$) &\ty 0.08   &\s 51458.967$^{\m e}$ &\s 4.11\ty($-0.16$) &\ty 0.07  &\s 50170.398$^{\m d}$ &\s 3.84\ty($-0.16$) &\ty 0.06   &\s 48910.831$^{\m a}$ &\s 3.50\ty($-0.20$) &\ty 0.40  &   \multicolumn{3}{c}{\s $\nu$=43 GHz}               \\           
                                                                                                                                                                                                                                              \cline{13-15}       
\s 50170.506$^{\m d}$ &\s 3.93\ty($-0.02$) &\ty 0.07   &\s 51458.991$^{\m e}$ &\s 4.46\ty($-0.11$) &\ty 0.07  &\s 50170.458$^{\m d}$ &\s 3.50\ty($-0.10$) &\ty 0.20   &\s 48910.982$^{\m a}$ &\s 4.40\ty($+0.10$) &\ty 0.40  &\s 49803.542$^{\m b}$ &\s 3.60\ty($-0.20$) &\ty 0.40 \\           
\s 50170.565$^{\m d}$ &\s 4.28\ty($-0.08$) &\ty 0.07   &\s 51459.019$^{\m e}$ &\s 4.59\ty($-0.10$) &\ty 0.06  &\s 50170.517$^{\m d}$ &\s 4.12\ty($-0.13$) &\ty 0.05   &\s 48911.134$^{\m a}$ &\s 3.00\ty($+1.00$) &\ty 0.40  &\s 49819.488$^{\m c}$ &\s 3.10\ty($+0.70$) &\ty 0.80 \\           
\s 50170.624$^{\m d}$ &\s 4.19\ty($-0.02$) &\ty 0.08   &\s 51459.046$^{\m e}$ &\s 4.95\ty($+0.03$) &\ty 0.06  &\s 50170.576$^{\m d}$ &\s 4.84\ty($-0.07$) &\ty 0.05   &\s 49803.385$^{\m b}$ &\s 2.50\ty($+0.10$) &\ty 0.30  &\s 49833.453$^{\m c}$ &\s 2.00\ty($+0.20$) &\ty 0.90 \\                                                                                                                                                                                                                   
\s 50170.684$^{\m d}$ &\s 3.78\ty($-0.06$) &\ty 0.08   &\s 51459.070$^{\m e}$ &\s 4.98\ty($-0.06$) &\ty 0.06  &\s 50170.636$^{\m d}$ &\s 4.69\ty($-0.06$) &\ty 0.06   &\s 49803.430$^{\m b}$ &\s 3.80\ty($+0.00$) &\ty 0.25   &\s 49837.400$^{\m c}$ &\s 3.10\ty($+0.90$) &\ty 0.90 \\                                                                                                                                                                                                                           
\s 50171.380$^{\m d}$ &\s 3.14\ty($+0.03$) &\ty 0.08   &\s 51459.094$^{\m e}$ &\s 4.60\ty($-0.15$) &\ty 0.07  &\s 50170.695$^{\m d}$ &\s 4.56\ty($-0.14$) &\ty 0.06   &\s 49803.475$^{\m b}$ &\s 4.30\ty($+0.60$) &\ty 0.25  &\s 53031.763$^{\m g}$ &\s 2.60\ty($+0.60$) &\ty 0.20 \\           
\s 50171.439$^{\m d}$ &\s 2.90\ty($+0.02$) &\ty 0.08   &\s 51459.117$^{\m e}$ &\s 4.97\ty($-0.08$) &\ty 0.07  &\s 50171.391$^{\m d}$ &\s 3.57\ty($-0.02$) &\ty 0.06   &\s 49803.535$^{\m b}$ &\s 4.20\ty($-0.10$) &\ty 0.30   &\s 53031.814$^{\m g}$ &\s 3.00\ty($+0.20$) &\ty 0.30 \\       
\s 50171.499$^{\m d}$ &\s 3.14\ty($-0.02$) &\ty 0.07   &\s 51459.141$^{\m e}$ &\s 4.52\ty($+0.04$) &\ty 0.07  &\s 50171.451$^{\m d}$ &\s 3.56\ty($+0.09$) &\ty 0.06   &\s 49803.580$^{\m b}$ &\s 3.40\ty($-0.20$) &\ty 0.25  &\s 53031.843$^{\m g}$ &\s 4.20\ty($+0.60$) &\ty 0.30 \\          
\s 50171.558$^{\m d}$ &\s 3.55\ty($+0.03$) &\ty 0.08   &\s 51459.163$^{\m e}$ &\s 3.60\ty($+0.00$) &\ty 0.10  &\s 50171.510$^{\m d}$ &\s 4.03\ty($-0.03$) &\ty 0.06   &\s 49803.624$^{\m b}$ &\s 3.20\ty($-0.30$) &\ty 0.25  &&& \\          
\s 50171.617$^{\m d}$ &\s 4.00\ty($-0.06$) &\ty 0.08   &\s 51461.935$^{\m e}$ &\s 4.18\ty($+0.20$) &\ty 0.06  &\s 50171.570$^{\m d}$ &\s 3.74\ty($-0.10$) &\ty 0.06   &\s 49820.603$^{\m c}$ &\s 6.00\ty($-0.60$) &\ty 0.30  &&& \\

\hline            
\end{tabular}     
\begin{list}{}{}
\item[] \s {Code: $^{\m a}$ AL267; $^{\m b}$ AL348; $^{\m c}$ AL346; $^{\m d}$ AL372; $^{\m e}$ AT233; $^{\m f}$ AL568; $^{\m g}$ AL618.}
\end{list}
\end{center}

\end{table*}

\section{Radio observations and data reduction}
Multi-frequency   observations of $\sigma$\,Ori\,E 
were carried out with the
VLA\footnote{The Very Large Array is a facility of the National Radio
Astronomy Observatory which is operated by Associated Universities, Inc.
under cooperative agreement with the National Science Foundation.} 
in different epochs.
Table~\ref{VLA_log} reports 
the instrumental and observational details
for each data set.

To avoid significative phase fluctuations, scans on $\sigma$\,Ori\,E were shorter than
the Earth atmosphere coherence time and embedded between phase calibrator measurements.
The  22 and 43 GHz observations were carried out using the {fast
switching mode} between source and phase calibrator.

The data were calibrated and mapped using the standard procedures
of the Astronomical Image Processing System (AIPS).
The flux density for the Stokes I parameter
was obtained by fitting a two-dimensional gaussian (JMFIT)
at the source position in the cleaned maps integrated over contiguous scans,
the integration time ranges from about 10 minutes to about 1 hour.
As the uncertainty in the flux density
measurements we assume the r.m.s. of the map.

The fraction of the circularly polarized flux density (Stokes V parameter)
was instead determined performing the direct Fourier transform
of the visibilities at the source position (DFTPL) { without any temporal average}.
We are justified in this procedure by the absence of other circularly polarized sources in the field.
{ Stokes V were later averaged with the same integration time of Stokes I.}
All VLA measurements are listed in Table~\ref{tab}.

\section{Radio properties of $\sigma$\,Ori\,E}
\subsection{Light curves}
Fig.~\ref{dati} shows the radio flux versus the
rotational phase for $\nu \leq 15$ GHz. For completeness, we plot
also the literature measurements at 5 GHz by \citet{drake_etal87} and \citet{leone_umana93}.
The top panel shows the variability of $B_{\rm e}$ \citep{oksala_etal10, landstreet_borra78, bohlender_etal87}.

As already known \citep{leone91,leone_umana93,trigilio_etal04}
the light curve at 5 GHz of $\sigma$~Ori~E is
characterized by the presence of two maxima
associated with the extrema of the magnetic field.
We observe a similar modulation in the flux curve at 1.4 GHz.
At both 1.4 and 5 GHz,
the minima in the light curves
are close in phase with the null magnetic field,
that occurs when the axis of the dipole is perpendicular to the line of sight.


The shape of the light curves at 8.4 and 15 GHz are more complex
and any relation with the $B_{\rm e}$ variability is no longer simple.
At 8.4 GHz more than two maxima are detected, two of these are clearly associated with the
two maxima observed at $\Phi \approx 0.3$ and 0.95 in the 15 GHz light-curve.
The amplitude of the  light curves increases with the frequency. It
ranges from about 
49\% (at 1.4 GHz) to about 82\% 
(at 15 GHz) of the median.

{ With the aim to detect CU\,Vir type pulses in  $\sigma$\,Ori\,E,
we focus our attention in the phase windows where we expect to see them.
Those windows are around the rotational phases corresponding to the null of the
magnetic field curve. The first null is at $\Phi\approx 0.9 - 0.1$, the second
at $\Phi\approx 0.3 - 0.5$, both with a duration $\Delta \Phi_\mathrm{null}\approx0.2$.
The phase coverage of the radio measurements around the first 
null is almost continuous in the frequency range 1.4 -- 15 GHz,
while around the second null there are fewer data.
Assuming a pulse extends over a phase window $\Delta \Phi_\mathrm{ECM} \approx 0.05$, as
for the narrowest pulse observed in CU\,Vir,
%
we estimate that the probability to have missed an ECM pulse
in the first null is zero, since the maximum phase gap $\Delta \Phi_\mathrm{gap}$ between 
contiguous points is smaller than $\Delta \Phi_\mathrm{ECM}$ for all the observed frequencies.
On the contrary, around the second null 
there are several gaps for which $\Delta \Phi_\mathrm{gap}>\Delta \Phi_\mathrm{ECM}$.
In this case, we estimate the probability to miss the ECM pulse as 
$(\Delta \Phi_\mathrm{gap}-\Delta \Phi_\mathrm{ECM})/\Delta \Phi_\mathrm{null}$.
%
Considering all the gaps in this phase window, those probabilities are  
63\%,  25\%, 16\% and 14\%,  respectively at 1.5, 5, 8.4 and 15 GHz.
For the whole rotational period, instead, we get 13\%, 5\%, 12\% and 15\%.
}

\begin{figure*}
\resizebox{\hsize}{!}{\includegraphics{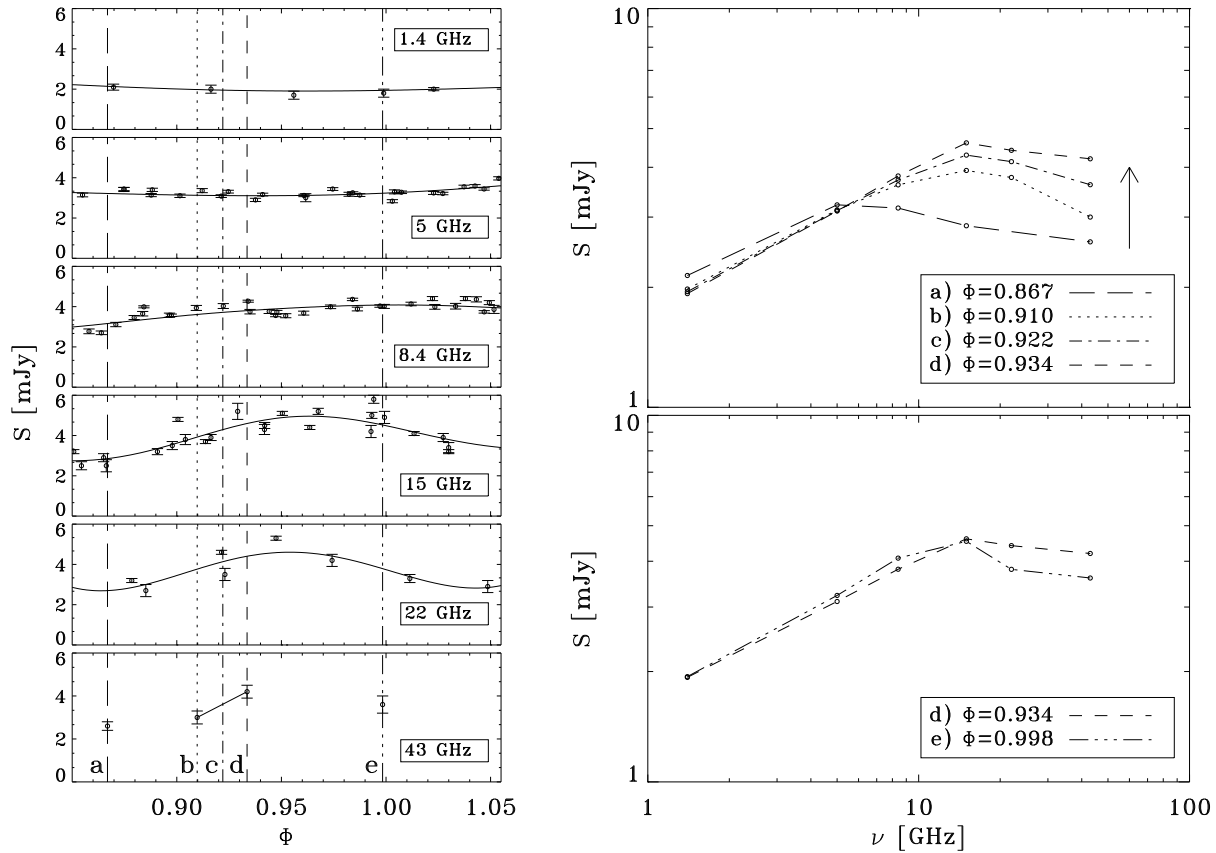}}
\caption{
Left panels: zoom of the radio light curves at phases where high frequencies
data are available; vertical lines (a, b, c, d, e) show the phases where the spectra have been realized.
Right panels: radio spectra at phases (a), (b), (c), (d), and (e); rising phase up panel,
decay phase down panel.
}
\label{spettri}
\end{figure*}

\subsection{Radio spectrum}
The 22 and 43 GHz VLA measurements  of $\sigma$\,Ori\,E 
do not cover the whole rotational period.
However, our 
22 and 43 GHz observations 
are in coincidence with the steep flux increment observed at 15 GHz ($\Phi \approx 0.95$).
We can then trace the complete radio spectrum from 1.4 to 43 GHz
in the range of phases: 0.9 -- 1, when the flux at 15 GHz approximately doubles. 
To minimize the effect of the lack of simultaneity of higher frequency measurements,
we have fixed the phases: $\Phi=$0.867, 0.910, 0.922, 0.934 and 0.998, coinciding with
the measurements at 43 GHz and with the interpolated value between the two closest in phases 43 GHz observations.
Fluxes at the other frequencies have been obtained by fits of the light curves,
see Fig.~\ref{spettri}, left panels.
The five spectra are labelled: (a), (b), (c), (d), and (e) in Fig.~\ref{spettri}.

The radio spectrum (a), obtained 
before the rising phase of the 15 GHz flux, 
is peaked at intermediate frequencies (5 -- 8.4 GHz). The
radio spectra ((b), (c) and (d)) just before the 15 GHz maximum 
are, instead, peaked at higher frequency (15 GHz) and 
show a steep rise at 43 GHz (Fig.~\ref{spettri}, right top panel).
At frequencies below 15 GHz the
spectral index is positive, with $\alpha\approx 0.3$ ($S_{\nu} \propto \nu^{\alpha}$).
The spectrum (e), obtained  after the maximum, 
closely resembles the spectrum (d) (having similar 15 GHz flux level but obtained before the maximum)
at the lowest frequencies,
whereas at the highest frequencies (22 and 43 GHz) fluxes are decreasing.

\subsection{Circular polarization}
The circularly polarized
radio emission shows two
extrema of opposite signs (Fig.~\ref{simulaz}, bottom panels), detected above the 3$\sigma$ detection threshold,
associated with the magnetic field extrema (Fig.~\ref{fit_b}).
The degree of polarization enhances as the frequency increases.
Positive degree of circular polarization is detected when the north magnetic pole is close
to the line of sight and negative in presence of the south pole.
When the magnetic poles are close to the direction of the line of sight
we observe most of the radially oriented field lines.
In this case the gyrosynchrotron mechanism give rise to
radio emission partially polarized, respectively right hand for the north pole
and left hand for the south pole.

The overall behavior of the circularly polarized radiation from
$\sigma$\,Ori\,E resembles so closely the case of CU\,Vir at 5, 8.4 and 15 GHz \citep{leto_etal06}.
This can be considered the typical behavior
of the gyrosynchrotron emission from a magnetosphere
characterized by a mainly dipolar symmetry.

\section{Effect of the magnetic quadrupole on the radio emission}
\subsection{Comparison with the 3D model}
In order to investigate  the radio emission from MCP stars,
in \citet{trigilio_etal04} we developed a 3D model to
compute the gyrosynchrotron emission from a magnetosphere
shaped in the framework of the oblique rotator model, that assumes a simple magnetic dipole.
\citet{leto_etal06}  extended the 3D model to solve the radiation
transfer for the circularly polarized emission (Stokes V parameter).
Such a 3D modeling has been successfully applied to $\sigma$\,Ori\,E and HD\,37017 by
\citet{trigilio_etal04} and to CU\,Vir by \citet{leto_etal06}. 

{
In this model a radiatively driven stellar wind flows along the magnetic field lines. 
The wind channeled by the magnetic field reaches the magnetic equator. 
At the Alfv\'en radius, where the kinetic energy density of the wind exceeds the magnetic one, the stellar wind breaks the magnetic field lines generating current sheets,
where particle acceleration occurs. The energetic electrons moving towards the stellar surface emit by gyrosynchrotron mechanism.
This thin layer is named middle magnetosphere 
and it is connected to the annular regions around the magnetic poles at high magnetic latitudes.
At lower magnetic latitudes the wind is instead confined in a dead zone (inner magnetosphere), 
resulting in an accumulation of circumstellar matter whose density decreases outward,
in accord with the magnetically confined wind shock model \citep{babel_montmerle97}.
This circumstellar material absorbs the continuous gyrosynchrotron radiation, 
resulting in the observed deep rotational modulation.
}

\begin{figure*}
\resizebox{\hsize}{!}{\includegraphics{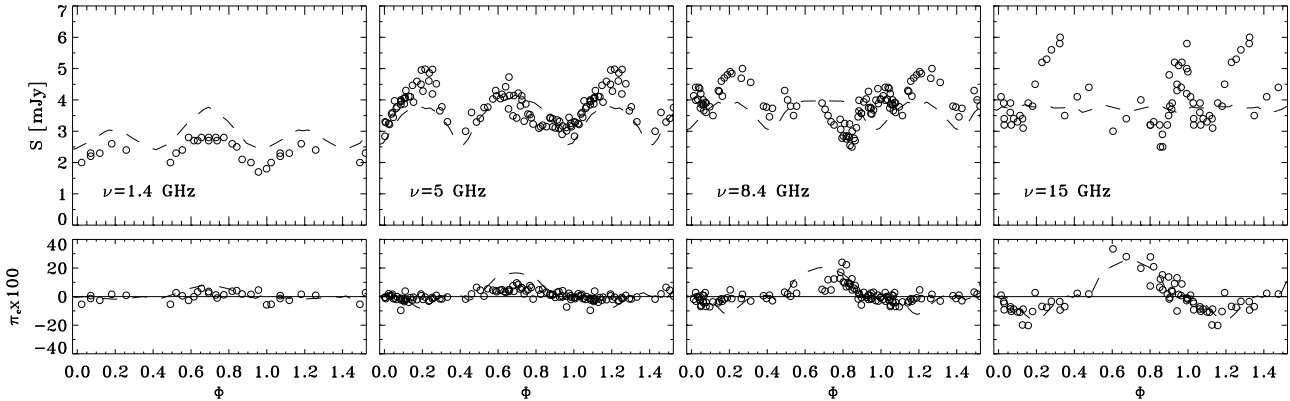}}
\caption{Comparison of the observed radio light-curves of $\sigma$\,Ori\,E and the theoretical curves (dashed line) computed with the 3D model by \citet{trigilio_etal04}. 
}
\label{simulaz}
\end{figure*}

Here we adopt the model parameters given by \citet{trigilio_etal04}, 
used to fit the literature 5 GHz light curve of  $\sigma$\,Ori\,E,
to compute the expected radio flux density and fraction of circular polarization
($\pi _{\mathrm c} = S_{\mathrm V}/S_{\mathrm I}$) for all the frequencies.
In Fig.~\ref{simulaz} we have compared  the simulations with the observations. 
The 1.4 GHz Stokes I variability is reproduced in shape, however slightly overestimating the average flux. 
As to the highest frequencies, our 3D model fails to reproduce the observed
Stokes I variability. On the contrary, Stokes V simulations are in good 
agreement with the measurements.

Since the source is
optically thick, the radiation at different frequencies
probes the stellar magnetosphere at different depths, where $\tau \approx 1$.
Higher frequencies are mainly emitted in the inner regions, where magnetic field is stronger.
In particular, inspecting the modeled 
radio maps of $\sigma$\,Ori\,E shown by \citet{trigilio_etal04}, the bulk of the 5 GHz
gyrosynchrotron emission arise from magnetospheric regions 3 -- 8 R$_{\ast}$
away from the stellar surface, whereas the 15 GHz originates from inner layers (1 -- 5 R$_{\ast}$).

We conclude that the 3D model reproduces the behavior of the 1.4 and 5~GHz radio emission,
coming mainly from regions far from the star, where the dipolar component of
the magnetic field dominates. 
The nearly-dipolar symmetry can be also identified from the 
behavior of the polarized emission,
well reproduced by the model simulations at all the frequencies (Fig.~\ref{simulaz}, bottom panels).
At 8.4 and 15~GHz the model instead fails to reproduce the observed Stokes I modulation,
as expected if a non dipolar magnetic field component is present.

In the case of gyrosynchrotron mechanism,
the radio emission from a region with almost parallel magnetic field lines directed toward us
(North), or in opposite direction (South), is right-hand (RCP), or left-hand (LCP), circularly polarized; Stokes V, 
which is the difference between RCP and LCP, is positive, or negative; 
Stokes I, which is the sum of RCP and LCP is sensitive to both polarizations.
In a region with dishomogeneous magnetic  field on a small scale, the polarization is, in average, null. 
When a quadrupolar field is present, we see always regions with antiparallel field lines; Stokes V is always 
null, while Stokes I depends on the strength of each individual component.
In this simple way we can explain the behavior of Stokes V, which well agrees with the dipolar model at any
frequency, and of the low frequency Stokes I, arising from regions where only the dipolar field is relevant. 
The high frequency Stokes I, instead, is sensitive to the quadrupolar component, important in the low magnetosphere.
We are supported in such a conclusion from the Stokes I and V variability 
presented at 5~GHz by HR\,5624 \citep*{lim_etal96}, that
is characterized by a multi-polar magnetic field \citep{landstreet90}.

The spectra shown in Fig.~\ref{spettri}  evidence as the radio flux density
at the higher frequencies steeply rise going up to the maximum at phase $\approx 0.95$, 
clearly observed in the 15 GHz light curve.
This could be 
a consequence of strong dishomogeneity in the magnetic field close to the star.

\subsection{The absence of the cyclotron maser}

This work is aimed mainly at searching for CU\,Vir-like pulses from $\sigma$\,Ori\,E.
In the following we will compute the expected frequency range of the ECM in our target.

\begin{figure}
\resizebox{\hsize}{!}{\includegraphics{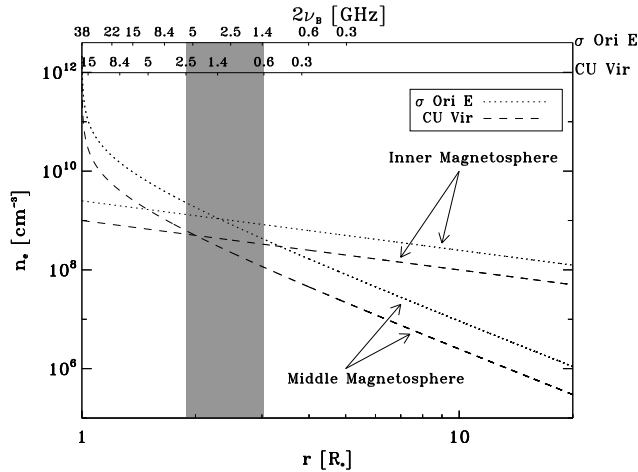}}
\caption{Radial dependence of the thermal electron density number 
inside the ``middle" and the ``inner" magnetosphere, in the cases 
of $\sigma$\,Ori\,E and CU\,Vir. 
In the top x-axis the ECM frequencies (2$\nu_\mathrm{B}$) for the two stars,
are shown, the shaded area identifies the corresponding magnetospheric regions.
} 
\label{fig5} 
\end{figure}

The minimum maser frequency observed in CU\,Vir is 610\,MHz \citep{stevens_george10},
{ the maximum is  2.5 GHz}.
Since the maser frequency is a harmonic of the local gyrofrequency, $\nu =s\times \nu_{\mathrm B}$,
with $s=2$, and assuming a simple dipolar field, for which the magnetic field strength above 
the pole is given by
\begin{displaymath}
B_{\rm{dip}}=B_{\m p} ({\m R}_{\ast}/R)^3,
\end{displaymath}
we can deduce that for CU\,Vir ($B_{\m p} = 3000$ G)  the maser at 610\,MHz is generated at  
$2~{\m R}_{\ast}$ above the stellar surface.
{ The shaded area of Fig.~\ref{fig5} defines the frequencies and the
corresponding distances associated with the ECM pulses as observed on CU\,Vir}. 
At the same height, for $\sigma$\,Ori\,E, with a polar field strength $B_{\m p} = 6800$~G,
the corresponding frequency of emission is 1.4 GHz.
Therefore the cyclotron maser emission from 
 $\sigma$\,Ori\,E, if it does exist, should be observable at frequencies higher than 1.4 GHz.
The maximum possible frequency is given by the second harmonic 
of the gyrofrequency very close to the stellar surface, corresponding to about 38 GHz.

Moreover, a necessary condition for the ECM is that the local plasma frequency  must be lower than the local 
gyrofrequency, i.e. $\nu_{\mathrm p}\ll \nu_{\mathrm B}$ \citep{melrose_dulk82}.
Since the plasma frequency is given by $\nu_{\mathrm p}=9\times 10^{-6} n_{\mathrm e} ^{1/2}$ GHz,
the maser emission near to the magnetic pole is possible only if the local thermal
electron number density is less than about $4\times 10^{12}$ cm$^{-3}$.
In the current model the thin cavity where
the maser should originate is the middle magnetosphere above the magnetic pole, where the local plasma is 
the ionized thermal wind that flows along the magnetic field lines. 
Following \citet{trigilio_etal08} and adopting the characteristic values of the wind derived with the 3D model
for $\sigma$\,Ori\,E \citep{trigilio_etal04}, we estimate that the density of the wind is about $10^{11}$ cm$^{-3}$ near the 
stellar surface, well below the theoretical quenching limit.
As the magnetic field strength decreases going outward, 
the plasma frequency equates the gyrofrequency at approximately 7\,R$_{\ast}$ above the surface.
Therefore, we can conclude that the condition for generation of the cyclotron maser 
is satisfied in the middle magnetosphere of $\sigma$\,Ori\,E.

{ The radial behavior of the plasma density in the middle and inner magnetosphere, is shown in Fig.~\ref{fig5} 
 for $\sigma$\,Ori\,E \citep{trigilio_etal04} and CU\,Vir \citep{leto_etal06}.
As recently observed on CU\,Vir \citep{trigilio_etal11,lo_etal12} 
the ray path of the ECM, which is polarized in the $x-$mode, is refracted by the plasma trapped 
in the inner magnetosphere, the refractive index is given by
$n_{\rm{refr}} \approx \sqrt{ 1 -  { \nu_{\rm{p}}^2 }/{ 2\nu_{\rm{B}}^2 } }$,
with $\nu_\mathrm{ECM}\approx2\nu_{\rm{B}}$.
Since the conditions of the plasma are quite similar for $\sigma$\,Ori\,E and CU\,Vir (Fig.~\ref{fig5}), 
we expect an upwards refraction. 
In few words the radiation should be deviated, not absorbed.}

Despite the light curves in the expected frequency range of the ECM (between 1.4 and 15 GHz) are well sampled, in particular near the rotational phases where we expect to detect it, i.e. in coincidence with 
at least one null of the magnetic curve (Fig.~\ref{dati}), we do not  see any hint of the pulses as observed on CU\,Vir.
Since the frequency of 1.4 and 15~GHz corresponds to the second harmonic of the local gyrofrequency
at 2 and 0.4 R$_{\ast}$ above the pole respectively, we can conclude that in this region 
the conditions needed to trigger the cyclotron maser are not fulfilled.

As highlighted in this paper,  the magnetic field of $\sigma$\,Ori\,E is characterized by the presence
of a quadrupolar  component, whose strength $B_{\rm{quad}}$, which is prevailing in the stellar surface,
decreases faster than the dipolar component  ($B_{\rm{quad}}\propto R^{-5}$, $B_{\rm{dip}}\propto R^{-3}$).
Therefore the actual magnetic vector in the low magnetosphere
can be quite different from the simple dipolar one.
As already discussed, this quadrupolar component significantly affects the gyrosynchrotron emission at 
high frequency, originating in the deep magnetospheric layers. In a similar way, the  quadrupolar component of the magnetic field, 
stronger then the dipolar one at low altitude,
affect the propagation of the electrons moving towards the stellar surface, causing the moving up of the
magnetic mirroring point and consequently different interaction with the thermal plasma of the stellar magnetosphere, and possible
changing of the ECM conditions.
In fact, for a charged particle moving with a pitch angle $\phi$ in a not uniform magnetic field, 
its magnetic moment is invariant, i.e.  $\sin ^2 \phi  / B=$ constant, 
leading to magnetic mirroring when the particle moves in a converging
magnetic field.
Assuming for $\sigma$\,Ori\,E a simple dipolar field, a non-thermal electron injected with 
$\phi _0 \approx 1^{\circ}$ at the Alfv\'en radius 
($R_{\rm {Alfven}}\approx 15$ R$_{\ast}$), where $B_{\rm{Alfven}} \approx 2$ G, 
will reach the stellar surface with a pitch angle $\phi \approx 90^{\circ}$ 
($\sin ^2 \phi =  \sin ^2 \phi _0  \times B_{\m p} / B_{\rm {Alfven}}$).
If there is a magnetic quadrupolar component, 
which is two or three times stronger than the strength of the dipolar field, 
the same electron is reflected at 0.3 -- 0.35 R$_{\ast}$ from the surface, and cannot reach the
deeper and denser magnetospheric layers; in this case it is reflected back without being absorbed.
Only those electrons injected at $R_{\rm {Alfven}}$ with a very small pitch angle ($\ll 1^{\circ}$) can
reach the stellar surface.
Accordingly, the fraction of non-thermal electrons that can develop an anisotropic pitch angle distribution is very small, and then the loss cone angle is already closed when the electrons reach the region where
the maser at frequency below 15 GHz should arise.
The quadrupole component vanishes 5 -- 6 R$_{\ast}$ away from the surface and
this explain the observed $\sigma$\,Ori\,E behavior. 


\section{Can O-type stars emit ECM pulses?}

{
Hot O-type stars have strong ionized stellar winds and, 
in few cases, organized magnetic fields 
\citep{donati_etal02,donati_etal06b}, like MCP stars. 
Such characteristics imply that the magnetic O-type stars may be sources of ECM.

Typically MCP stars have weak winds 
($\dot{M} \approx 10^{-10} - 10^{-9}$ M$_{\odot}$ yr$^{-1}$)
strongly dominated by the magnetic fields ($B_{\mathrm p} \approx$ 1--10 kG), with large co-rotating 
magnetospheres.
The measured magnetic field in the O-type stars is about 1 kG or less \citep{hubrig_etal11}
and the mass loss rate is higher then $10^{-7}$ M$_{\odot}$ yr$^{-1}$.
Since $R_{\mathrm {Alfven}} \propto B_{\mathrm p}^{1/2} /  \dot{M}^{1/4}$ \citep{uddoula_etal09},
the very massive wind and the relatively weak magnetic field locates the Alfv\'en surface of the 
known magnetic O-type stars 
close to the stellar surface ($< 2$ R$_{\ast}$).
In the case of single magnetic O-type stars, shocks originating in the wind channeled by the magnetic field 
can accelerate electrons up to relativistic energy
\citep*{vanloo_etal06}.
These, in turn, could emit at the radio regime by gyrosynchrotron mechanism. 
On the other hand the O-type stars are characterized by radio emission at centimeter wavelengths 
mainly ascribed to the thermal free-free emission from the ionized stellar wind,
whose typical spectral index is $\alpha \approx 0.6$ ($S_\nu\propto\nu^{\alpha}$) \citep{wright_barlow75}.
The radio photosphere, at centimeter wavelengths,  has a radius of $\approx 100$~R$_{\ast}$ \citep{blomme11},
meaning that the plasma of the wind is optically thick inside and that any non-thermal emission generated
within is completely absorbed \citep{vanloo_etal06}.
The ECM from O-type stars, if any, would suffer from the same absorption effect.

Moreover, for a main sequence magnetic O-type star with stellar radius $R_{\ast}=7-8$ R$_{\odot}$,
radiative wind mass loss rate $\dot{M} = 10^{-6}$ M$_{\odot}$ yr$^{-1}$,
terminal velocity $v_{\infty} = 2000$ km s$^{-1}$ and dipolar magnetic field with
$B_{\mathrm p}=1000$~G, the condition 
$\nu_\mathrm{p} > \nu_\mathrm{B}$ is valid everywhere.
This is a further consideration against the possibility that a magnetic O-type
star can emit ECM pulse.
Up to now the signatures of non-thermal emission mechanism, flat or negative spectral index and variability,
in the O-type stars 
are observed only in multiple systems, the relativistic electrons origin is thus ascribed to 
the acceleration in the shocks that take place in the colliding winds  far from the stellar surface \citep{blomme11}.

In conclusion, the conditions for generation and propagation
of the ECM emission are not satisfied for the known magnetic O-type stars.
The behavior of the magnetized O-type stars is thus significantly different from 
the MCP stars (B/A spectral type) in the radio domain.
The key to explain when the transition occurs
is probably the shaping of the stellar magnetosphere,
which depends on
mass loss rate, magnetic field strength and rotational period.
}

\section{Conclusions and Outlook}
CU\,Vir is up to now the only known magnetic chemically peculiar star presenting
electron cyclotron maser emission.
The characteristics of the MCP star $\sigma$\,Ori\,E suggest that it 
should present the same
phenomenology of CU\,Vir at frequencies higher than 1.4 GHz. 
We have obtained  multi frequencies radio observations
during the whole the rotation period of  $\sigma$\,Ori\,E,
in particular at the phases where the ECM is expected.
However, no cyclotron maser emission has been detected in this star.

There are indications 
that the magnetic field topology of $\sigma$\,Ori\,E cannot be considered as a simple dipole.
The presence of the quadrupole field could inhibit the development of the conditions
able to power the cyclotron maser emission.
Nevertheless, we cannot rule out the possibility that the ECM could occur at 
{ other 
frequencies not analyzed in this paper, such as the 2.5 GHz,
or at phases near the second null of the magnetic curve
that are not fully covered.}

On the other hand, to better understand the ECM in the wider context of plasma processes
it will be extremely interesting to extend such investigation to a bigger sample of MCPs.
The average radio luminosity of the MCP stars is about $10^{16.8 \pm 0.9}$ [ergs s$^{-1}$ Hz$^{-1}$]
\citep{drake_etal87,linsky_etal92};
it has been also shown that 
the radio luminosity of the MCP stars increases with the effective temperature
\citep{linsky_etal92,leone_etal94}.
The radio interferometers of new generation, like the EVLA or the forthcoming 
Australian SKA Pathfinder (ASKAP, which will operate at 1.4 GHz),
will allow to reach the detection limit of few $\mu$Jy in all sky deep surveys.
Assuming a threshold of 10 $\mu$Jy
we estimate in about 2000 pc the maximum distance within will
be possible to detect the radio emission from the MCPs.
Following \citet{renson_manfroid09}
we can assume the MCP stars uniformly distributed in space.
It is reasonable to expect that the number of radio detections will increase of about an order of magnitude, 
giving the opportunity to get a larger statistics of the physical conditions of the magnetospheres, like
the magnetic field strength and orientation and the thermal plasma density, to correlate with the ECM.

We want to stress here that this class of objects provides the unique possibility to study plasma process in 
stable magnetic structures, whose topologies are quite often  well determined by several independent diagnostics
\citep{bychkov_etal05}, thus overcoming the variability of the magnetic field  that is one of the major problem that prevents  an accurate modeling of ECM in very active stars such dMe or close binary systems. 
{ On the other side, moving towards earlier spectral types, stronger and denser stellar winds 
together with a weaker magnetic field  
should inhibit the onset of the ECM instability.} 
Moreover, as shown for the prototype CU\,Vir, the observations of persistent coherent pulses of ECM, 
which act as a clock of the star, in other MCP stars would offer a valuable tool for the angular momentum evolution of the objects belonging to this class.

\section*{Acknowledgments}

We thank the referee for his/her constructive criticism which enabled us to improve this paper.


\end{document}